\documentclass[twocolumn,prb,showpacs,floatfix,superscriptaddress]{revtex4}
\usepackage[dvips]{color,graphicx}
\usepackage{amsfonts}
\usepackage{amssymb,amsmath,amsthm}
\usepackage{bm}

\newcommand{\be}{\begin{equation}}
\newcommand{\ee}{\end{equation}}
\newcommand{\bea}{\begin{eqnarray}}
\newcommand{\eea}{\end{eqnarray}}

\newcommand{\jc}{\hat{J}}

\begin{document}

\title{Drude weight in systems with open boundary conditions}
\author{Marcos Rigol}
\affiliation{Department of Physics, University of California, Santa Cruz, California 95064, USA}
\affiliation{Department of Physics, Georgetown University, Washington, District of Columbia 20057, USA}
\author{B. Sriram Shastry}
\affiliation{Department of Physics, University of California, Santa Cruz, California 95064, USA}

\date{\today}
\pacs{05.60.Gg,72.10.Bg}
%05.60.Gg Quantum transport
%72.10.-d Theory of electronic transport; scattering mechanisms
%72.10.Bg General formulation of transport theory

%--------------------------------------------------------------------------
\begin{abstract}
For finite systems, the real part of the conductivity is usually decomposed as the sum 
of a zero frequency delta peak and a finite frequency regular part. In studies with 
periodic boundary conditions, the Drude weight, i.e., the weight of the zero frequency 
delta peak, is found to be nonzero for integrable systems, even at very high temperatures, 
whereas it vanishes for generic (nonintegrable) systems. Paradoxically, for systems with 
open boundary conditions, it can be shown that the coefficient of the zero frequency delta 
peak is identically zero for any finite system, regardless of its integrability. In order 
for the Drude weight to be a thermodynamically meaningful quantity, both kinds of boundary 
conditions should produce the same answer in the thermodynamic limit. We shed light on 
these issues by using analytical and numerical methods. 
\end{abstract}
%--------------------------------------------------------------------------

\maketitle
%**************************************************************************
% Introduction
%**************************************************************************

Transport properties define materials as superconductors, metals, or 
insulators.\cite{kohn64,shastry90,scalapino92} In one dimension, transport 
can help differentiate integrable from nonintegrable systems. 
Integrable systems, in general, display an infinite conductivity 
at all temperatures,\cite{castella95} an effect that has been related 
to the strong influence of conservation laws, while at high 
temperatures, generic (nonintegrable) systems are expected to exhibit a finite 
resistivity caused by umklapp scattering and inelastic collisions.

Let us consider a finite system of length $L$ (in units of the lattice constant); 
the real part of the conductivity\footnote{For concreteness, we restrict our analysis to the 
electrical conductivity. Our conclusions are equally relevant to the thermal 
conductivity due to electrons.} can be written as\cite{castella95,shastry90,shastry06}
\bea
\textrm{Re} \left[ \sigma_L(\omega)\right] & = & \pi D_L \delta(\hbar \omega)  
+ \frac{\pi}{L} \left( \frac{1 - e^{ -\beta \hbar \omega}}{\omega}\right) 
\nonumber \\
&&\times\sum_{\epsilon_n \neq \epsilon_m} p_n  |J_{nm}|^2  
\delta( \epsilon_m- \epsilon_n - \hbar \omega),
\label{conductivity}
\eea
where
\bea
D_L= \frac{1}{L} \left[  \langle \hat{\Gamma} \rangle - \hbar \sum_{\epsilon_n \neq \epsilon_m} 
\frac{ p_n- p_m }{\epsilon_m - \epsilon_n} |J_{nm}|^2  \right], 
\label{drude1}
\eea
is the Drude weight (or charge stiffness), the stress tensor operator
$\hat{\Gamma}=-q_e\lim_{k \rightarrow 0} \frac{1}{k} [ \jc(k),\hat{n}(-k)]$, $\jc$ 
is the current operator, $J_{nm}=\langle n|  \jc | m  \rangle$ are its matrix elements, 
$\hat{n}(k)$ is the Fourier transform of the local density operator $\hat{n}_j$, $q_e$ 
the charge, and $\Gamma=\langle\hat{\Gamma}\rangle/\hbar L$. The Boltzmann weight 
of a state $| m \rangle$ is denoted by $p_m=e^{-\beta \epsilon_m}/Z$, with $\epsilon_m$ 
being its energy, $\beta=1/k_BT$, and $Z$ the partition function.

In one dimension, and finite temperatures, the Drude weight can also be computed 
as,\cite{castella95,shastry06} (assuming that there is no true superconductivity)
\bea
D^*_L= \frac{\beta \hbar}{L} \sum_{\epsilon_n=\epsilon_m} p_n |J_{nm}|^2.
\label{drude2}
\eea

Equation (\ref{conductivity}) is true for all boundary conditions. In the thermodynamic 
limit (TL), it leads to the decomposition $\textrm{Re} \left[ \sigma_{\infty}(\omega)\right] 
= \pi D_ \infty\delta(\hbar \omega) + \sigma_{reg}(\omega)$, where the infinite volume object
$D_ \infty=\lim_{L\rightarrow \infty} D_L=\lim_{L\rightarrow \infty} D^*_L$, and $\sigma_{reg}$ 
is the ``regular'' part of the conductivity. In systems with periodic boundary conditions 
(pbc's), $D_L$ and $D^*_L$ are found to be nonzero in integrable systems at all temperatures, 
i.e., they display an infinite conductivity, whereas $D_L$ and $D^*_L$ are expected to exponentially vanish with the system size for nonintegrable systems.\cite{castella95,fabian03}

A very different situation arises when one studies systems with open boundary conditions 
(obc's). There, one can legitimately define\cite{shastry06} the position operator 
$\hat{X}=\sum_{j} j\, \hat{n}_j$, which satisfies the following
commutation relations: $\jc=i[\hat{X},\hat{H}]$, where $\hat{H}$ is the Hamiltonian
of the system and $\hat{\Gamma}=-i[\hat{X},\jc]$. Since 
$J_{nm}= i (\epsilon_m - \epsilon_n) \langle n | \hat{X} | m \rangle$,
one can see that $D^*_L$ (\ref{drude2}) is identically zero, and that 
the first and second terms in the right hand side of Eq.\ (\ref{drude1}) cancel each 
other, so that $D_L$ is also identically zero. The vanishing of $D_L$ and $D^*_L$ for 
systems with obc's is true regardless of their integrability.\footnote{It is known 
that the notion of integrability survives in most cases when pbc's are replaced by obc's, 
using reflected Yang-Baxter equations, see, e.g.,
C. M. Yung and M. T. Batchelor, Nucl. Phys. B {\bf 435}, 430 (1995);
E. K. Sklyanin, J. Phys. A {\bf 21}, 2375 (1988).}

Considering the results above, one could question whether the Drude weight is a meaningful 
thermodynamic quantity, i.e., whether one would obtain the same result, measuring it in 
(quasi-) one-dimensional (1D) experiments involving closed rings or open chains. This is a valid concern since $D_ \infty$ is not a usual bulk thermodynamic quantity. (For the latter, the 
equivalence of boundary conditions in the TL is obvious.) In an expansion of the energy 
of the system in powers of $L$, $D_ \infty$ at $T=0$ is proportional to a term that scales 
like $1/L$.\cite{kohn64,shastry90} Our goal in this work is to shed light on this puzzle.

We consider 1D systems of spinless fermions,
\bea
\hat{H}&=&-t\sum_j \hat{c}^\dagger_j \hat{c}_{j+1} + \textrm{h.c.} 
          +V\sum_j \hat{n}_j \hat{n}_{j+1} \nonumber \\
       && -t'\sum_j\hat{c}^\dagger_j \hat{c}_{j+2} + \textrm{h.c.}
          +V'\sum_j\hat{n}_j \hat{n}_{j+2}
\label{hamiltonian}
\eea
with nearest ($t$) and next-nearest ($t'$) neighbor hopping, interacting with nearest ($V$) and next-nearest ($V'$) neighbor repulsive potentials. The sum over $j$ is appropriately defined for pbc's and obc's. $c^\dagger_j$ ($c_j$) are the creation (annihilation) operators for spinless fermions at a given site $j$, and $\hat{n}_j=c^\dagger_jc_j$ are the corresponding density 
operators. This model is known to be integrable for $t'=V'=0$. (It can be mapped onto the 
{\it XXZ} spin 1/2 chain.) To understand the effects of obc's in more generic nonintegrable 
systems, we study systems with finite values of $t'$ and $V'$.

The current and stress tensor operators for this model can be written as
\bea
\jc&=&\frac{iq_et}{\hbar}   \sum_j \hat{c}^\dagger_j \hat{c}_{j+1} + \textrm{H.c.}
   +\frac{2iq_et'}{\hbar} \sum_j \hat{c}^\dagger_j \hat{c}_{j+2} + \textrm{H.c.}, \nonumber \\
\hat{\Gamma}&=& \frac{q^2_et}{\hbar}   \sum_j \hat{c}^\dagger_j \hat{c}_{j+1} + \textrm{H.c.}
   +\frac{2q^2_et'}{\hbar} \sum_j \hat{c}^\dagger_j \hat{c}_{j+2} + \textrm{H.c.}. \nonumber
\eea

As a starting point for our analysis, let us consider the case of noninteracting fermions.
With pbc's, the single-particle eigenstates of the Hamiltonian are plane waves 
$| m \rangle =1/\sqrt{L}\sum_j e^{ik_mj}\hat{c}^\dagger_j |0\rangle$ with energy 
$\epsilon_m=-2t\,\cos(k_m)$, where $k_m=2\pi m/L$ and $m=-L/2+1,\ldots,L/2$. In this case, 
$|J_{nm}|^2=[2q_et\,\sin(k_m)]^2\delta_{m,n}$, i.e., the matrix elements of $\jc$ are fully diagonal, and $D_L=\langle\hat{\Gamma}\rangle/\hbar L=q^2_e\langle\hat{H}\rangle/\hbar L$. 
At finite temperatures, and for sufficiently large system sizes, $D_L^*=D_L$.

For obc's, on the other hand, standing waves being  the noninteracting eigenstates,  have the form 
$| m \rangle =\sqrt{2/(L+1)}\sum_j\sin(k_mj)\hat{c}^\dagger_j |0\rangle$
with energy $\epsilon_m=-2t\,\cos(k_m)$, where $k_m=\pi m/(L+1)$ and $m=1,\ldots,L$.
The matrix elements of the current operator are then
\[
|J_{nm}|^2 = \frac{8q^2_et^2}{(L+1)^2}\left[ 1- (-1)^{(m-n)}\right] 
\frac{\sin^2(k_m) \sin^2(k_n)}{\left[ \cos(k_m) - \cos(k_n)\right]^2}, 
\]
and vanish whenever $(m-n)$ is an even number, i.e., only eigenstates with a 
different parity produce nonvanishing values of $J_{nm}$. This expression shows
that the matrix elements of $\jc$ have a very interesting property: although 
$|J_{nm}|$ vanish for $m=n$, they attain their largest values 
for the smallest (odd) differences between $m$ and $n$. For very large 
systems, the factor $1/L^2$ produces vanishing values of $J_{nm}$ 
unless $n\sim m$. Rewriting $k_n=k_m\pm \pi l/(L+1)$, $l=1,3,\ldots$ one can 
see that $\textrm{Re} \left[ \sigma(\omega)\right]$ is a sum of delta functions 
$\delta [\omega \pm 2\pi t\,\sin(k_m) l/(L+1)]$ whose weight decreases as
$\sim 1/l^2$.

%%%%%%%%%%%%%%  FIGURE  %%%%%%%%%%%%%%%%%%%%%%%%%%%%%%%%%%%%%%%%%%%%%%
\begin{figure}[!tb]
\begin{center}
  \includegraphics[width=0.44\textwidth,angle=0]{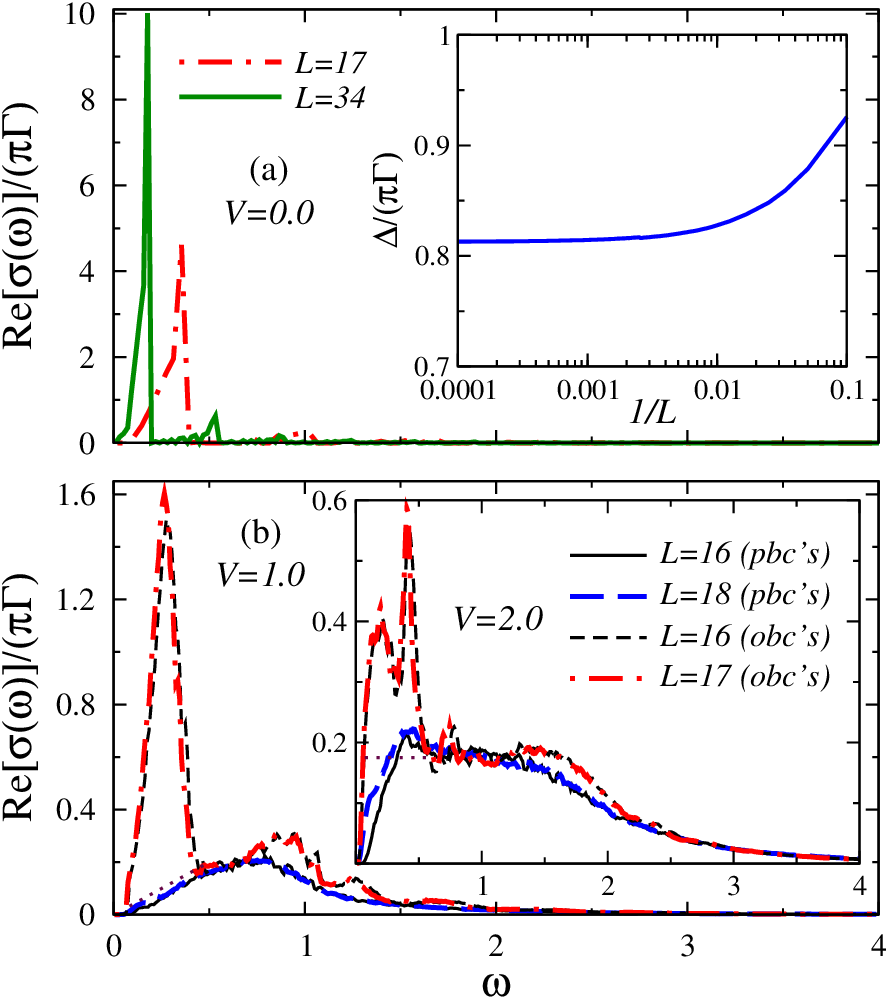}
\end{center}
\vspace{-0.6cm}
\caption{\label{Conductivityvsw_I} (color online). Regular part of 
$\textrm{Re} \left[ \sigma(\omega)\right]$ for integrable systems ($t'=V'=0$)
in (a) the noninteracting case ($V=0$), (b) $V=1.0$, and [inset in (b)] the 
isotropic (Heisenberg) point $V=2.0$. All of these results were obtained at a 
very high temperature $\beta=0.001$ (infinite in the noninteracting case) 
and half-filling. The inset in (a) shows how $\Delta$ (see text) behaves as one 
approaches the TL. The values of $V$, $\beta$, and $\omega$ are 
given in units of $t$.}
\end{figure}
%%%%%%%%%%%%%%%%%%%%%%%%%%%%%%%%%%%%%%%%%%%%%%%%%%%%%%%%%%%%%%%%%%%%%%%%

In Fig.\ \ref{Conductivityvsw_I}(a), we show $\textrm{Re} \left[ \sigma(\omega)\right]$
(binned to obtain a smooth curve) for noninteracting particles in two systems with obc's 
and different sizes. This figure shows that (i) broadened delta functions 
collapse to larger peaks (whose width reduces as the system size increases) 
situated at $\omega \sim 2\pi l/L$ with $l=1,3,\ldots$, and (ii) 
these peaks move toward smaller frequencies as $L$ increases.
From the previous analysis, one can conclude that although for finite 
open systems $D_L$ and $D^*_L$ are always zero, for $L\rightarrow \infty$, a delta 
peak develops at $\omega=0$, but this time generated by the collapse of 
delta peaks that come from the so-called regular part of the conductivity. 
In addition, from the sum rule  
$\int_{-\infty}^\infty \textrm{Re}\left[ \sigma(\omega)\right]  d\omega = { \pi}\langle 
\hat{\Gamma} \rangle/\hbar L$,\cite{shastry06} one obtains that the weight 
of such $\omega=0$ 
peak is $\pi D_\infty = \lim_{L\rightarrow \infty}\pi \langle\hat{\Gamma}\rangle/\hbar L
=\lim_{L\rightarrow \infty}\pi q^2_e \langle\hat{H}\rangle/\hbar L$, 
which is identical to $\pi D_\infty$ as obtained from periodic 
systems since the energy is identical in both cases.

At this point, one may wonder about the behavior of the finite frequency
($\omega \sim \pm 2\pi l/L$, $l=1,3,\ldots$) peaks as $L$ is increased. 
It may happen that as $L\rightarrow \infty$, (i) all the weight is concentrated 
in the lowest $\omega = \pm 1/L$ peaks, and the others disappear, or (ii) the weight
is distributed among several peaks with different values of $\omega$. To answer this 
question, we have studied the summed weight of the peaks at $\omega \sim \pm 2\pi/L$ 
as a function of increasing system size, a quantity we call $\Delta$. Results 
are shown in the inset in Fig.\ \ref{Conductivityvsw_I}(a). They confirm the scenario (ii)
above since, as $L\rightarrow \infty$, $\Delta$ saturates at around 80\% of $\pi \Gamma$. 
Hence, the other peaks with $l>1$ remain finite, and they are needed to account for the 
exact Drude weight in the TL.

Moving away from the noninteracting case, but keeping the system integrable, we cannot make 
the corresponding analytical treatment, so we turn to a full exact diagonalization of finite chains. 
We perform calculations in the grand-canonical ensemble, and all results presented in what follows 
are obtained at half-filling ($n=0.5$). 

\begin{figure}[!tb]
\begin{center}
  \includegraphics[width=0.44\textwidth,angle=0]{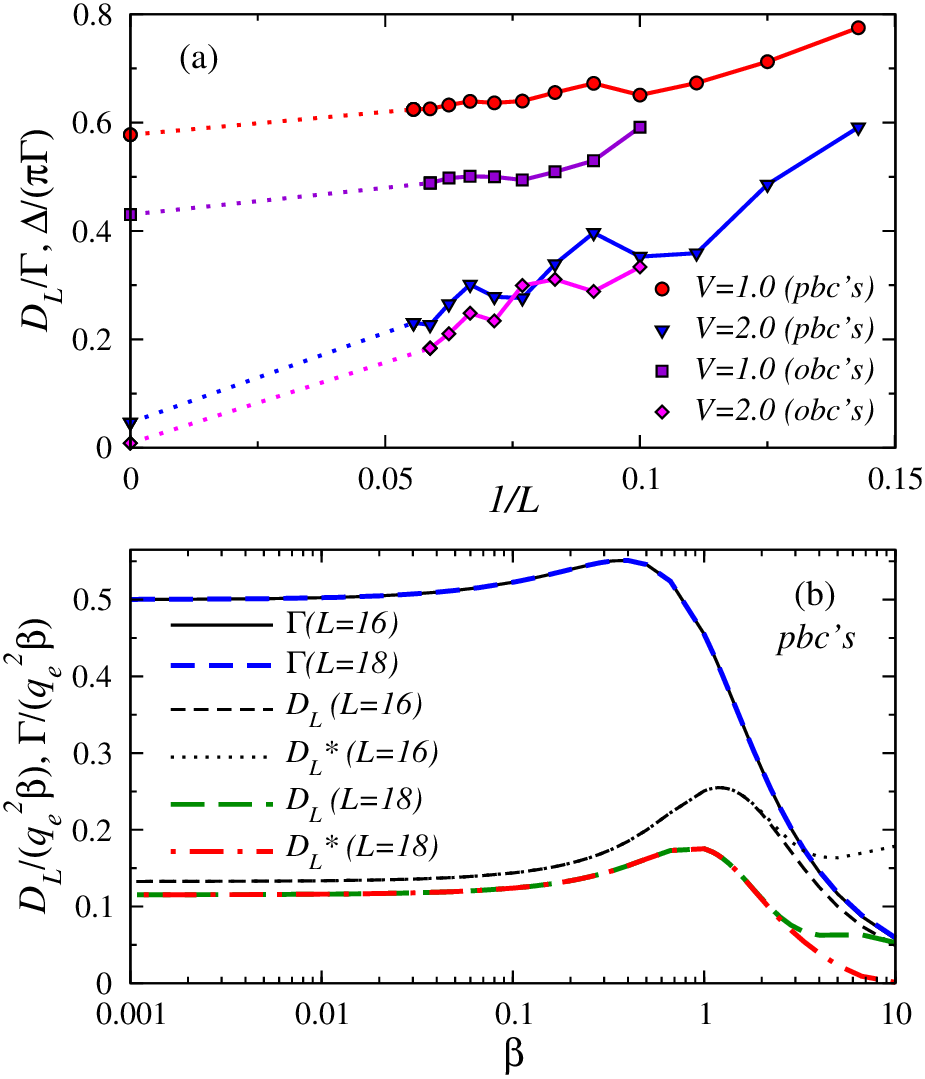}
\end{center}
\vspace{-0.6cm}
\caption{\label{DrudeWeight} (color online). (a) Scaling of the Drude weight 
(pbc's) and $\Delta$ (obc's) with 
the system size, for $V=1.0$ and $2.0$ and $\beta=0.001$. (b) The Drude weight, 
computed as $D_L$ and $D^*_L$, and $\Gamma$ as a function of $\beta$ for the 
isotropic case ($V=2.0$).}
\end{figure}

For generic integrable systems with pbc's, the matrix elements 
of $\jc$ are not diagonal anymore like in the noninteracting case. This means that 
weight moves away from the Drude peak and the regular part of 
$\textrm{Re} \left[ \sigma(\omega)\right]$ becomes finite. This can be seen 
in  Fig.\ \ref{Conductivityvsw_I}(b) and its inset, where we have plotted 
$\textrm{Re} \left[ \sigma(\omega)\right]$ for two integrable systems with different 
values of $V$. The scaling of the Drude weight with system size, together 
with a ``simple minded'' linear extrapolation to the TL, are shown in 
Fig.\ \ref{DrudeWeight}(a). As seen there, for both $V=1.0$ and $V=2.0$, we obtain 
a finite value of $D_\infty$.\cite{shastry90,castella95,fabian03,mukerjee07}

%%%%%%%%%%%%%%  FIGURE  %%%%%%%%%%%%%%%%%%%%%%%%%%%%%%%%%%%%%%%%%%%%%%
%%%%%%%%%%%%%%%%%%%%%%%%%%%%%%%%%%%%%%%%%%%%%%%%%%%%%%%%%%%%%%%%%%%%%%%%

The results presented so far were obtained at a very high temperature ($\beta=0.001$), 
where finite size effects are the smallest, but transport properties are still nontrivial. 
In Fig.\ \ref{DrudeWeight}(b), one can see that for $T>10$, the actual value of $T$ is not
essential since quantities such as $D_LT$, $D^*_LT$, and $\Gamma T$ become almost independent 
of $T$. Figure \ref{DrudeWeight}(b) also shows that, for the considered 
temperatures and system sizes, $\Gamma$ is independent of the system size, 
while $D_L$ and $D^*_L$ still exhibit finite size effects. It is important to notice 
that albeit $D_L$ and $D^*_L$ are identical for any given system size at high 
temperatures, finite size effects build differences between these two ways of computing
the Drude weight at lower temperatures.\cite{fabian03}

After reviewing the pbc case, we can now analyze the effects of obc's on more generic
integrable systems. In Fig.\ \ref{Conductivityvsw_I}(b) and its inset, we have plotted 
$\textrm{Re} \left[ \sigma(\omega)\right]$ for chains with obc's, and two different values 
of $V$, together with the results for pbc's. One can clearly see there that large peaks 
develop in the obc data over the pbc results, and that these peaks move toward 
lower frequencies as $L$ is increased.\cite{kuhner00} Actually, for $V=1.0$, one can 
see that the two largest obc peaks are located at $\omega \sim 2\pi l/L$, $l=1$ and 3, 
similar to the noninteracting case. These results strongly suggest that in the TL, the 
finite frequency peaks for obc's will collapse into a single $\omega=0$ Drude peak like 
the one obtained for pbc's.

%%%%%%%%%%%%%%  FIGURE  %%%%%%%%%%%%%%%%%%%%%%%%%%%%%%%%%%%%%%%%%%%%%%
\begin{figure}[!tb]
\begin{center}
  \includegraphics[width=0.44\textwidth,angle=0]{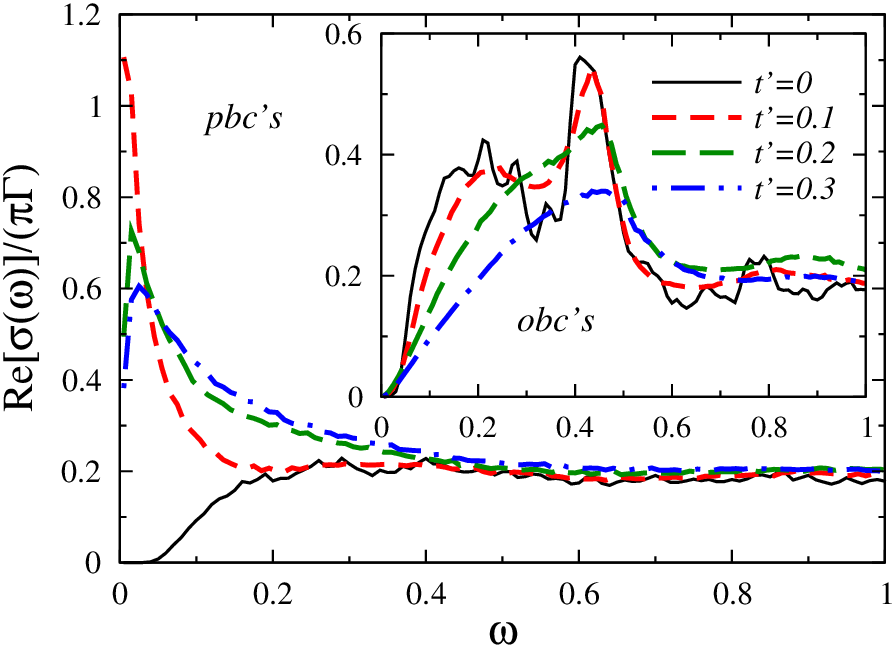}
\end{center}
\vspace{-0.6cm}
\caption{\label{Conductivityvsw_tp01to03} (color online). Regular part of 
$\textrm{Re} \left[ \sigma(\omega)\right]$ for nonintegrable systems 
with pbc's ($L=17$) and (inset) obc's ($L=16$) as one moves away from the isotropic 
integrable point by increasing the value of $t'$. In all cases, $V=2.0$ 
and $\beta=0.001$.}
\end{figure}
%%%%%%%%%%%%%%%%%%%%%%%%%%%%%%%%%%%%%%%%%%%%%%%%%%%%%%%%%%%%%%%%%%%%%%%%

In Fig.\ \ref{DrudeWeight}(a), we show how $\Delta$ behaves with increasing system size
for $V=1.0$ and $V=2.0$. [$\Delta$ is computed in this case as two times the area over 
the dotted lines in Fig.\ \ref{Conductivityvsw_I}(b).] In both cases, the behavior of 
$\Delta$ is consistent with extrapolated $\omega=0$ peaks with a finite weight in the 
TL. Like  for the noninteracting case, extrapolating $\Delta$ to 
$L\rightarrow \infty$ does not reproduce the value of $D_\infty$ obtained for 
systems with pbc's. This, we infer, is due to the weight distributed over peaks 
with higher frequencies, which, following the noninteracting case, should 
all collapse to $\omega=0$ as $L\rightarrow \infty$. The extrapolations in 
Fig.\ \ref{DrudeWeight}(a) show that the relative weight of peaks with higher 
$\omega$ in chains with obc's increases as one departs from the noninteracting case.

Having shed light on integrable systems, we now turn to the nonintegrable case. 
As mentioned before, in systems with pbc's, breaking integrability is expected to produce 
an exponentially (with the system size) vanishing Drude weight. We first consider 
(Fig.\ \ref{Conductivityvsw_tp01to03}) the case in which in Eq.\ (\ref{hamiltonian}) 
integrability is broken by $t'$, which also breaks the particle-hole symmetry of the 
integrable model. In Fig.\ \ref{Conductivityvsw_tp01to03}, one can see 
that, even for small finite systems, introducing $t'$ has dramatic effects 
in $\textrm{Re} \left[ \sigma(\omega)\right]$. A peak develops at low frequencies. 
Given the sum rule for $\textrm{Re} \left[ \sigma(\omega)\right]$, such a peak can be 
related to the disappearance of the Drude peak and the transfer of its weight to finite 
frequencies. Hence, the system acquires a finite dc conductivity, which decreases with 
increasing $t'$ (Fig.\ \ref{Conductivityvsw_tp01to03}). 

For small finite systems with obc's, on the other hand, nothing dramatic should happen 
when a small $t'$ is introduced. This is because in the integrable case, there is no zero 
frequency delta peak, but, instead, finite frequency peaks that are already present 
in the regular part of $\textrm{Re} \left[ \sigma(\omega)\right]$. As seen in the 
inset in Fig.\ \ref{Conductivityvsw_tp01to03}, adding $t'$ to integrable chains with 
obc's only reduces the height of the lowest frequency peak in 
$\textrm{Re} \left[ \sigma(\omega)\right]$ and, with increasing $t'$, its weight 
moves toward higher frequencies. 

%%%%%%%%%%%%%%  FIGURE  %%%%%%%%%%%%%%%%%%%%%%%%%%%%%%%%%%%%%%%%%%%%%%
\begin{figure}[!tb]
\begin{center}
  \includegraphics[width=0.45\textwidth,angle=0]{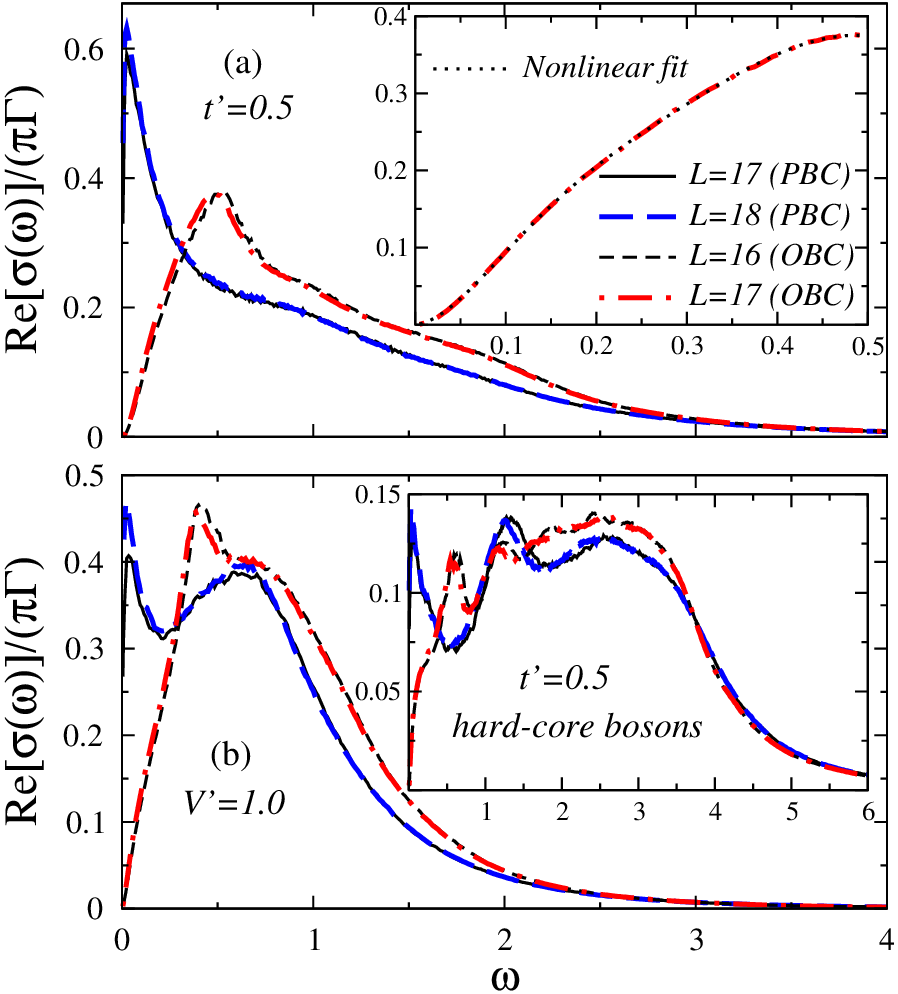}
\end{center}
\vspace{-0.6cm}
\caption{\label{Conductivityvsw_NI} (color online). 
Regular part of $\textrm{Re} \left[ \sigma(\omega)\right]$ 
for nonintegrable chains with different system sizes. (a) $t$-$t'$-$V$ model with $t'=0.5$,
(b) $t$-$V$-$V'$ model with $V'=1.0$, and [inset in (b)] $t$-$t'$-$V$ model for a system of 
hard-core bosons with $t'=0.5$. The inset in (a) shows the low-frequency part of 
$\textrm{Re} \left[ \sigma(\omega)\right]$ for the fermionic $t$-$t'$-$V$ model 
together with a nonlinear fit (see text). In all cases, $V=2.0$, $\beta=0.001$, 
and $t'$, $V$, $V'$, $\beta$, and $\omega$ are given in units of $t$.}
\end{figure}
%%%%%%%%%%%%%%%%%%%%%%%%%%%%%%%%%%%%%%%%%%%%%%%%%%%%%%%%%%%%%%%%%%%%%%%%

An apparent difference between $\textrm{Re} \left[ \sigma(\omega)\right]$ in systems with 
pbc's and systems with obc's (Fig.\ \ref{Conductivityvsw_tp01to03}) is that, while the former 
exhibit a finite dc conductivity, the latter exhibit a vanishing one. The vanishing of 
the dc conductivity for finite systems with obc's is understandable 
by using an analogy with pbc's with a small but finite momentum $q$. For $q\neq 0$, we know
from the conservation law [$\omega\, \hat{n}(q,\omega)\simeq q\,\hat{J}(q,\omega)$]
that the current must vanish in the dc limit.\cite{mukerjee07} One expects, 
in this case, to see diffusion, i.e., $\sigma(q,\omega)=\sigma(0,\omega)/[1-i D q^2/\omega]$ 
with $q\sim \pi/L$ for obc's. This means that at very low frequencies for obc's,
$\textrm{Re} \left[ \sigma(\omega)\right]\sim \omega^2$. As shown in the inset in 
Fig.\ \ref{Conductivityvsw_NI}(a), our data are crudely consistent with that. There, 
we also show the result of a nonlinear fit to $\textrm{Re} \left[ \sigma(\omega)\right]=
\sum_{a=1,6} C_a \omega^a$, where we find that $C_1\simeq 0$.

From the previous analysis, one expects the region of $\omega$'s over which 
$\textrm{Re} \left[ \sigma(\omega)\right] \rightarrow 0$ to reduce with increasing
$L$ and, eventually, in the TL, to recover the $k=0$ result
obtained for pbc's. In the main panel of Fig.\ \ref{Conductivityvsw_NI}(a), we depict 
results for pbc's and obc's in systems with two different sizes. There, one can see 
that, indeed, for obc's, the low-frequency region with decreasing conductivity moves 
to lower frequencies with increasing system size. In the TL, the 
usual belief is that long time tails in the autocorrelation of the current will 
ultimately take over, asymptotically leading to 
$\textrm{Re} \left[ \sigma(\omega)\right] \sim A-B\sqrt{|\omega|}$.\cite{mukerjee06}

We should stress that our conclusions above are valid for generic nonintegrable systems, i.e., they are not limited to the $t$-$t'$-$V$ model presented in Figs.\ \ref{Conductivityvsw_tp01to03} and \ref{Conductivityvsw_NI}(a). For example, in Fig.\ \ref{Conductivityvsw_NI}(b) we show that similar results are obtained when $t'$=0 but $V'$$\neq$$0$ ($t$-$V$-$V'$ model, main panel), and $V'$=0, $t'$$\neq$$0$ ($t$-$t'$-$V$ model), but for a system of hard-core bosons (inset). The last two models preserve the particle-hole symmetry present in the integrable case, so our conclusions are also independent of its presence or absence in nonintegrable systems.

In summary, we have argued that even though the real part of the conductivity  in finite systems
with obc's is qualitatively different from that of systems with pbc's, they
both have a common thermodynamic limit. For integrable systems, this means that there is an infinite 
dc conductivity, characterized by a finite Drude weight. On the other hand, for nonintegrable 
systems, the dc conductivity is finite and, in our 1D systems, it decreases as one moves away
from the integrable point.

\vspace{-0.4cm}

\begin{acknowledgments}

\vspace{-0.2cm}

We acknowledge support from NSF-DMR-0706128 and DOE-BES DE-FG02-06ER46319. We thank F. Heidrich-Meisner and O. Narayan for stimulating discussions.

\end{acknowledgments}

\end{document}